\documentclass[preprint,12pt]{elsarticle}

\usepackage{amssymb}
\usepackage{amsmath}
\usepackage{xcolor}

\journal{Physics Letter A}

\begin{document}

\begin{frontmatter}


\title{Experimental verification of Generalised Synchronization\tnoteref{}}
\author{Tania Ghosh\corref{cor1}\fnref{label1}}
\ead{ghoshtania96@gmail.com}
\fntext[cor1]{Corresponding author}
\affiliation[label1]{
    organization={University of Houston}, 
    city={Houston}, 
    postcode={77204}, 
    state={Texas}, 
    country={USA}
}

 \author{Soumitro Banerjee\fnref{label2}}
\affiliation[label2]{
organization={Indian Institute of Science Education \& Research}, 
    addressline={Kolkata}, 
    state={West Bengal}, 
    postcode={741246}, 
     country={India}
}
\ead{soumitro@iiserkol.ac.in}

\begin{abstract}
Generalized synchronization (GS) describes a state in which two coupled dynamical systems exhibit a functional relationship between their variables. GS can be achieved by appropriately designing the coupling to constrain the dynamics onto an invariant submanifold, a concept well established in theory. However, experimental validation remains crucial. In this work, we experimentally demonstrate GS using coupled chaotic Lorenz oscillators, implementing unidirectional coupling where the state variables of the slave oscillator are scaled according to a predefined relationship with those of the master oscillator.
\end{abstract}



\begin{highlights}
\item We experimentally demonstrate generalized synchronization using coupled chaotic Lorenz oscillators, implementing unidirectional coupling where the state variables of the slave oscillator are scaled according to a predefined functional relationship with those of the master oscillator.
\end{highlights}


\end{frontmatter}


\section{Introduction}

Chaotic dynamical systems have been experimentally explored through electronic circuits~\cite{Kennedy_1992}, which serve as practical platforms to test theoretical models like the Lorenz~\cite{lorenz1963} and Rössler~\cite{Rossler_1976} oscillators. These circuits not only realize abstract dynamics physically but also validate their robustness in real-world settings. 

Synchronization of chaotic systems is a fundamental concept in nonlinear dynamics~\cite{Pecora_1990,Fujisaka_1983,Pecora_1997,Duolan_2023,CHEN_2005,LU_2004,Lin_2016,Irimiciuc_2015}, where coupled system trajectories align onto a hyperplane in phase space which is called the synchronization manifold~\cite{Pecora_1997,Sun2009}. If the manifold is transversally attracting, synchronization remains stable. This can be seen as constraining the system's flow to a lower-dimensional submanifold. 

Generalized synchronization (GS)~\cite{Rulkov_1995,KITTEL_1998} extends this concept, where subsystems exhibit a functional relationship governed by their coupling but are not identical under simple transformations like inversion or time delay. The functional relationship can be projective synchronization, nonlinear scaling, or any other form as long as it is smooth and invertible. Chishti and Ramaswamy \cite{Ramaswamy_2018} further generalized this to arbitrary submanifolds. 

Generalized synchronization (GS) has been experimentally demonstrated on various systems~\cite{KITTEL_1998,Uchida_2003,Abarbanel_1996,Parlitz_1996} spanning different frequency regimes, from high~\cite{Dmitriev_2009} to low frequencies~\cite{Uchida_2003}. In this Letter, we experimentally verify that coupling two electronic circuits with a targeted functional dependence between their variables can produce a specific state of generalized synchronization. In our study, we used two identical Lorenz oscillators, although the approach is applicable to any pair of nonlinear systems. We used two types of functional dependencies, projective synchronization and a non-linear scaling relationship, to verify generalized synchronization in the coupled systems.

\label{sec1}

\section{Theory of Generalized Synchronization}
\label{subsec1}

Consider two coupled dynamical systems:
\begin{equation}
\dot{x} = F_1(x) + \zeta(x, y), \quad \dot{y} = F_2(y) + \zeta(x, y),
\end{equation}
where \( x \in \mathbb{R}^m \) and \( y \in \mathbb{R}^n \) are the state variables and \( \zeta(x, y) \) are the coupling terms. Combining these into a vector \( X = \begin{bmatrix} x \\ y \end{bmatrix} \), the system becomes:
\begin{equation}
\dot{X} = F(X) + \epsilon(X) 
\end{equation}

We define a set of constraints \( \Phi(x, y) = 0 \) to constrain the dynamics to a submanifold \( M \) \cite{Ramaswamy_2018}. The flow must be orthogonal to the normal vectors of this submanifold, given by:
\begin{equation}
N_i(X) = \nabla_X \phi_i(x, y).
\end{equation}

\noindent The normal matrix is:
\begin{equation}
n = \nabla_X^\top \Phi(X) = \begin{bmatrix} N_1 & N_2 & \dots & N_N \end{bmatrix}^\top.
\end{equation}

\noindent The condition for invariance on the submanifold is:
\begin{equation}
 n \zeta = -n F,
\end{equation}
from which the coupling functions \( \zeta_i \) can be determined.   This methodology allows us to design couplings that ensure generalized synchronization, as introduced in \cite{Ramaswamy_2018}.


\section{Modified Projective Synchrony}

Among generalized synchronization (GS), a specific type of synchronization is Modified Projective Synchronization (MPS)~\cite{Koronovskii2014,Barajas_2012}, where the driving and response systems synchronize up to a scaling factor. 
Using the concept of MPS~\cite{Ramaswamy_2018}, the target submanifold in a six-dimensional phase space is considered as:
\begin{equation}
    y = \Lambda x
\end{equation}
where
\begin{equation}
    \Lambda =
    \begin{bmatrix}
        a & 0 & 0 \\
        0 & b & 0 \\
        0 & 0 & c
    \end{bmatrix}.
\end{equation}

As a specific example, we consider a system of two coupled Lorenz oscillators~\cite{lorenz1963}. Each oscillator has a flow equation in $\textbf{X} \in \textbf{$R^3$}$\\
\noindent\begin{minipage}{.5\linewidth}
\begin{equation}
  \renewcommand{\arraystretch}{1.2}
  \left.\begin{array}{r@{\;}l}
    \dot{x_{\rm 1}} &= \sigma(x_{\rm 1}-x_{\rm 2}) \\
    \dot{x_{\rm 2}} &=  x_{\rm 1}(\rho - x_{\rm 3}) - x_{\rm 2} \\
    \dot{x_{\rm 3}} &= x_{\rm 1}x_{\rm 2} - \beta x_{\rm 3}
  \end{array}\right\} \label{eq}
\end{equation}
\end{minipage}%
\begin{minipage}{0.5\linewidth}
\begin{equation}
  \renewcommand{\arraystretch}{1.2}
  \left.\begin{array}{r@{\;}l}
    \dot{y_{\rm 1}} &= \sigma(y_{\rm 1}-y_{\rm 2}) \\
    \dot{y_{\rm 2}} &=  y_{\rm 1}(\rho - y_{\rm 3}) - y_{\rm 2} \\
    \dot{y_{\rm 3}} &= y_{\rm 1}y_{\rm 2} - \beta y_{\rm 3}
  \end{array}\right\} \label{eq}
\end{equation}
\end{minipage}

\medskip
\noindent where $\sigma$, $\rho$, and $ \beta $ are their parameters.

Using the above method, \cite{Ramaswamy_2018} showed that one solution of unidirectional coupling can be obtained by setting $\zeta_1$ as a null vector and choosing $\zeta_2$ as  
\begin{equation}
\zeta_2 = 
\begin{bmatrix}
\sigma(a-b)x_2\\
(b-a)\rho x_1+(ac-b)x_1x_3\\
(c-ab)x_1x_2
\end{bmatrix}
\end{equation} 

\subsection{Circuit Implementation}

In this work, we implement the above system with unidirectional coupling, for which the master and slave equations are modified as follows:
\begin{equation}
\begin{aligned}
    \dot{x_1} &= \sigma(x_1 - x_2) & \quad \dot{y_1} &= \sigma(y_1 - y_2) + \sigma(a - b)x_2 \\
    \dot{x_2} &= x_1(\rho - x_3) - x_2 & \quad \dot{y_2} &= y_1(\rho - y_3) - y_2 + (b - a)\rho x_1 + (ac - b)x_1x_3 \\
    \dot{x_3} &= x_1x_2 - \beta x_3 & \quad \dot{y_3} &= y_1y_2 - \beta y_3 + (c - ab)x_1x_2
\end{aligned}
\label{eqn:unidirectional}
\end{equation}

The Lorenz equations were scaled down by a factor of 10 so that the variables remain within the useful range of electronic devices.

\begin{figure}[t]
\centering
\includegraphics[scale=0.5]{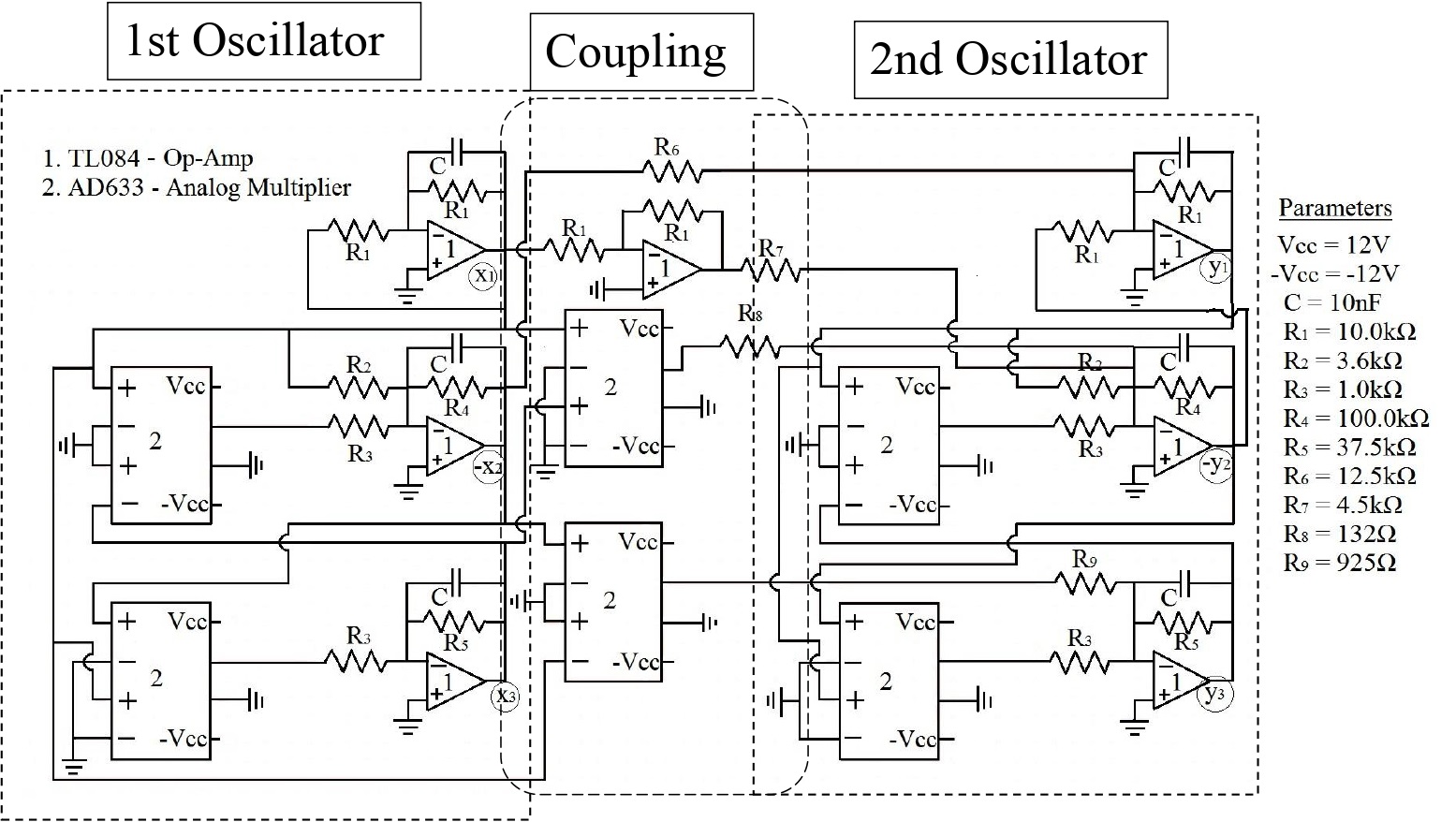}
\caption{Circuit implementation for projective synchronization, where \(y_1 = 2.6x_1\), \(y_2 = 1.8x_2\), and \(y_3 = 3.6x_3\).}
\label{fig:ckt1}
\end{figure}

Using the same technique used in~\cite{Kennedy_1992}, we transformed Eq.~\ref{eqn:unidirectional} into the final equations for implementation:
\begin{equation}
  \renewcommand{\arraystretch}{1.2}
  \left.\begin{array}{r@{\;}l}
    CR\dot{x_{\rm 1}} &= -\frac{R}{R_1}(x_{\rm 1}-x_{\rm 2}) \\
    CR\dot{x_{\rm 2}} &=-(\frac{R}{R_4} x_{\rm 2}-\frac{R}{R_2} x_{\rm 1} + \frac{R}{R_3} \frac{x_{\rm 1} x_{\rm 3}}{10}  ) \\
    CR\dot{x_{\rm 3}} &=-(\frac{R}{R_3} \frac{x_{\rm 1}x_{\rm 2}}{10}  - \frac{R}{R_5} x_{\rm 3})
  \end{array}\right\} \label{ckt_eqtna}
\end{equation}
\begin{equation}
  \renewcommand{\arraystretch}{1.2}
  \left.\begin{array}{r@{\;}l}
    CR\dot{y_{\rm 1}} &= -(\frac{R}{R_1}(y_{\rm 1}-y_{\rm 2})- \frac{R}{R_6} x_{\rm 2})\\
    CR\dot{y_{\rm 2}} &=-(\frac{R}{R_4} y_{\rm 2}-\frac{R}{R_2} y_{\rm 1} + \frac{R}{R_3} \frac{y_{\rm 1} y_{\rm 3}}{10}  + \frac{R}{R_7} x_{\rm 1}- \frac{R}{R_8}\frac{x_{\rm 1} x_{\rm 3}}{10} ) \\
    CR\dot{y_{\rm 3}} &=-(\frac{R}{R_3} \frac{y_{\rm 1}y_{\rm 2}}{10}  - \frac{R}{R_5} y_{\rm 3}+\frac{R}{R_9}\frac{x_{\rm 1}x_{\rm 2}}{10} )
  \end{array}\right\} \label{ckt_eqtnb}
\end{equation} 
We built an electronic circuit implementing Eq.~\ref{ckt_eqtna} and Eq.~\ref{ckt_eqtnb}, ensuring that the governing equations of the circuit align with those of Eq.~\ref{eqn:unidirectional}.
The circuit was constructed with AD633JN multipliers, TL084 operational amplifiers, resistors, and polyester capacitors (10 nF). The resistance values were selected as shown in Fig.~\ref{fig:ckt1} to achieve the scaling factors $a = 2.6$, $b = 1.8$, and $c = 3.6$. Fig.~\ref{fig:ckt1} provides a detailed schematic of the circuit, which incorporates unidirectional coupling.

\subsection{Experimental Result}

\begin{figure}[tbh]
\centering
\includegraphics[scale=.35]{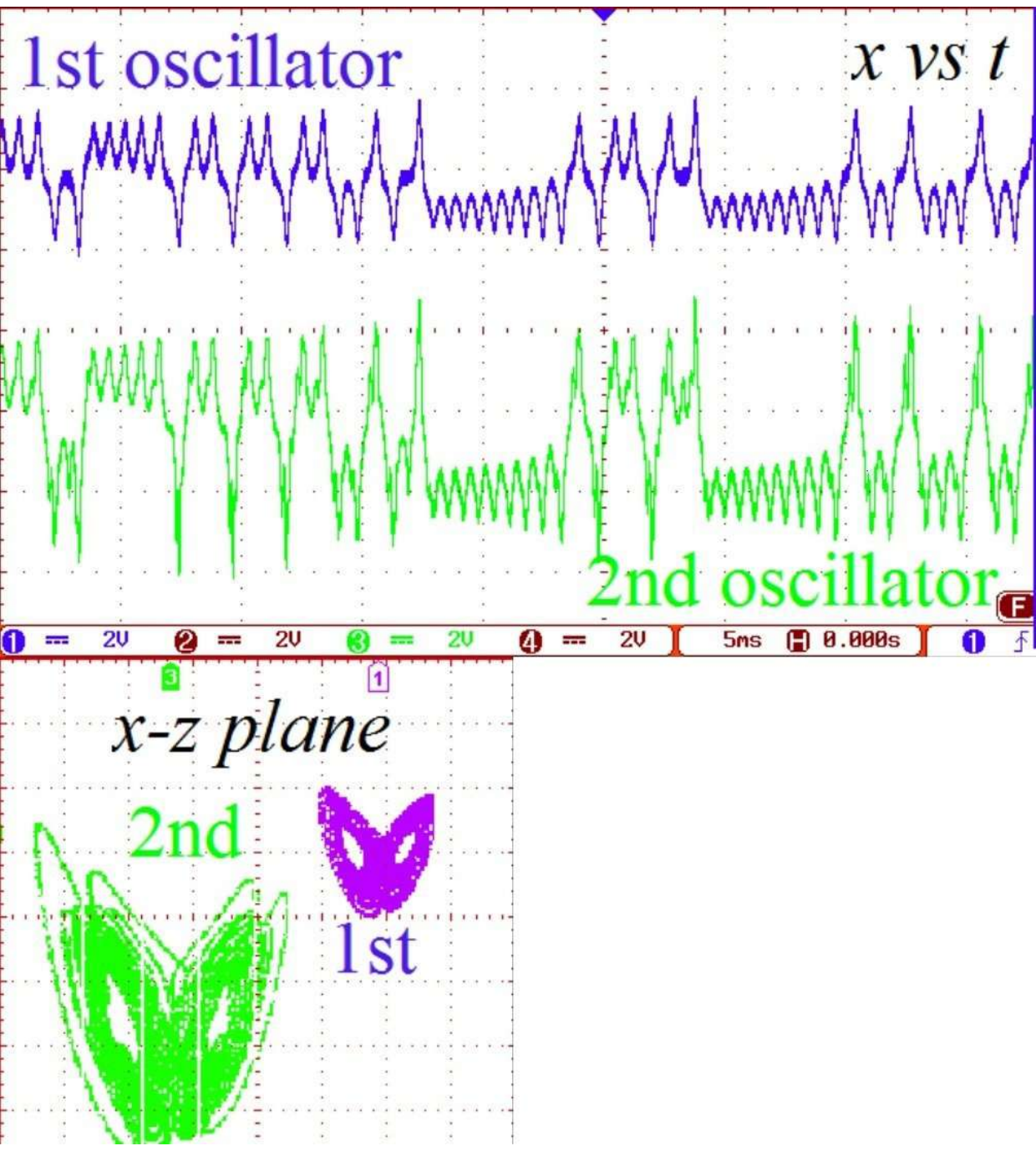}
\caption{Experimental results: synchronized time series and phase space trajectories. Projective synchronization in coupled Lorenz oscillators for the specific electronic circuit shown in Fig.~\ref{fig:ckt1}, using resistor values that result in scaling factors $a = 1.6$, $b = 1.8$, and $c = 3.6$. \label{fig:result1}}
\end{figure}

Fig.~\ref{fig:result1} demonstrates successful projective synchronization (PS) between two coupled Lorenz oscillators, with the relationships $
y_1 = 2.6 x_1, y_2 = 1.8 x_2$,
$y_3 = 3.6 x_3$.
In the figure, the signals of \(x_1\) and \(y_1\) with the \(x-z\) phase space for the two oscillators are shown as observed from the oscilloscope. The blue/purple traces represent the first (master) oscillator, and the green trace corresponds to the second (slave) oscillator. This confirms the achievement of generalized synchronization (GS), where the variables of the second oscillator exhibit a specific functional dependence on the first, expressed as $
y(t) = f(x(t))$,
with distinct scaling factors for each variable.

\section{Nonlinear Scaling}

The second example we consider involves the case where variable in the slave oscillator becomes a nonlinear function of a variable in the master oscillator. In particular, we intend to achieve the relationship \[ y_1 = x_1,\;\;  y_2 = x_2,\;\; \mbox{and}\;\; y_3 = \frac{x_3^2}{10}  + \frac{9x_3 }{10} \] for two coupled Lorenz systems.

For unidirectional coupling, one solution is \( \zeta_1 \) as a null vector, and \( \zeta_2 \) is given by:

\begin{equation}
\mbox{\boldmath$\zeta$}_2 = 
\begin{bmatrix}
0 \\
-\frac{x_3 y_1}{10} + \frac{x_1 x_3^2}{10} \\
-\frac{x_1 x_2}{10} + \frac{2 x_1 x_2 x_3}{10} - \frac{\beta x_3^2}{10}
\end{bmatrix}
\end{equation}

\medskip
\subsection{Circuit Implementation}

For nonlinear scaling with unidirectional coupling, the slave equation is modified as follows:
\begin{equation}
\begin{aligned}
    \dot{y_1} &= \sigma(y_1 - y_2) + 0 \\
    \dot{y_2} &= y_1(\rho - y_3) - y_2 -\frac{x_3 y_1}{10} + \frac{x_1 x_3^2}{10} \\
    \dot{y_3} &= y_1 y_2 - \beta y_3 - \frac{x_1 x_2}{10} + \frac{2 x_1 x_2 x_3}{10} - \frac{\beta x_3^2}{10}
\end{aligned}
\end{equation}

Equations for implementation are

\begin{equation}
  \renewcommand{\arraystretch}{1.2}
  \left.\begin{array}{r@{\;}l}
    CR\dot{x_{\rm 1}} &= -\frac{R}{R_1}(x_{\rm 1}-x_{\rm 2}) \\
    CR\dot{x_{\rm 2}} &=-(\frac{R}{R_4} x_{\rm 2}-\frac{R}{R_2} x_{\rm 1} + \frac{R}{R_3} \frac{x_{\rm 1} x_{\rm 3}}{10} ) \\
    CR\dot{x_{\rm 3}} &=-(\frac{R}{R_6} \frac{x_{\rm 1}x_{\rm 2}}{10}  - \frac{R}{R_5} x_{\rm 3})
  \end{array}\right\} \label{ckt_eqtnc}
\end{equation}
\bigskip

\begin{equation}
  \renewcommand{\arraystretch}{1.2}
  \left.\begin{array}{r@{\;}l}
    CR\dot{y_{\rm 1}} &= -\frac{R}{R_1}(y_{\rm 1}-y_{\rm 2})\\
    CR\dot{y_{\rm 2}} &=-(\frac{R}{R_4} y_{\rm 2}-\frac{R}{R_2} y_{\rm 1} + \frac{R}{R_3} \frac{y_{\rm 1} y_{\rm 3}}{10}  -\frac{R}{R_7}\frac{x_{\rm 1}x_{\rm 3}^2}{100} +\frac{R}{R_8}\frac{x_{\rm 1} x_{\rm 3}}{10} ) \\
    CR\dot{y_{\rm 3}} &=-(\frac{R}{R_5} y_{\rm 3}-\frac{R}{R_6} \frac{y_{\rm 1}y_{\rm 2}}{10}  -\frac{R}{R_9}\frac{x_{\rm 1}x_{\rm 2}x_{\rm 3}}{100} +\frac{R}{R_{10}}\frac{x_{\rm 1}x_{\rm 2}}{10}+ \frac{R}{R_{11}}\frac{x_{\rm 3}^2}{10} )
  \end{array}\right\} \label{ckt_eqtnd}
\end{equation}

\begin{figure}[tbh]
\centering
\includegraphics[scale=.85]{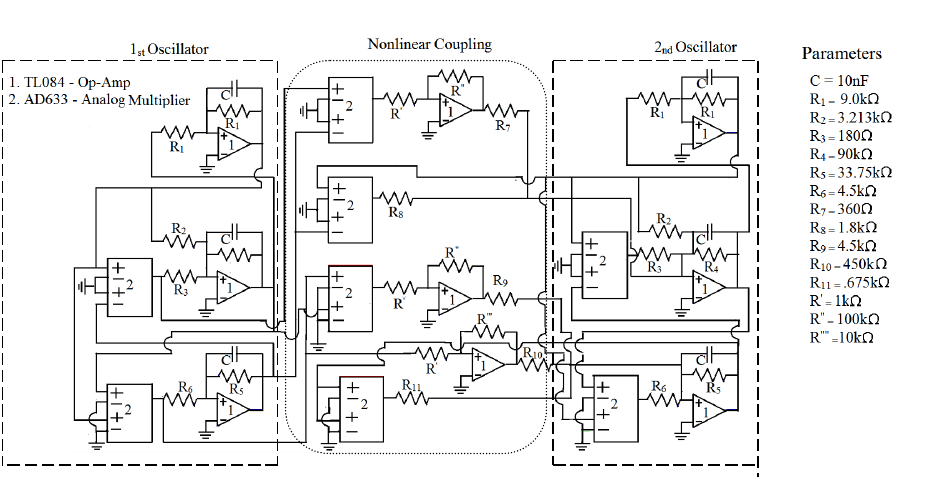}
\caption{Circuit implementation for nonlinear scaling, where \( y_1 = x_1 \), \( y_2 = x_2 \), and \( y_3 = \frac{x_3^2}{10}  + \frac{9x_3 }{10} \) }
\label{fig:ckt2}
\end{figure}
The circuit diagram for the nonlinear scaling, where $y_1 = x_1$, $y_2 = x_2$, $y_3 =  \frac{x_3^2}{10}+\frac{9x_3}{10}$,
is shown in Fig.~\ref{fig:ckt2}. The values of the resistors and capacitors used in the circuit are provided in the figure.

\subsection{Experimental Result}

\begin{figure}[tbh]
\centering
\includegraphics[scale=.4]{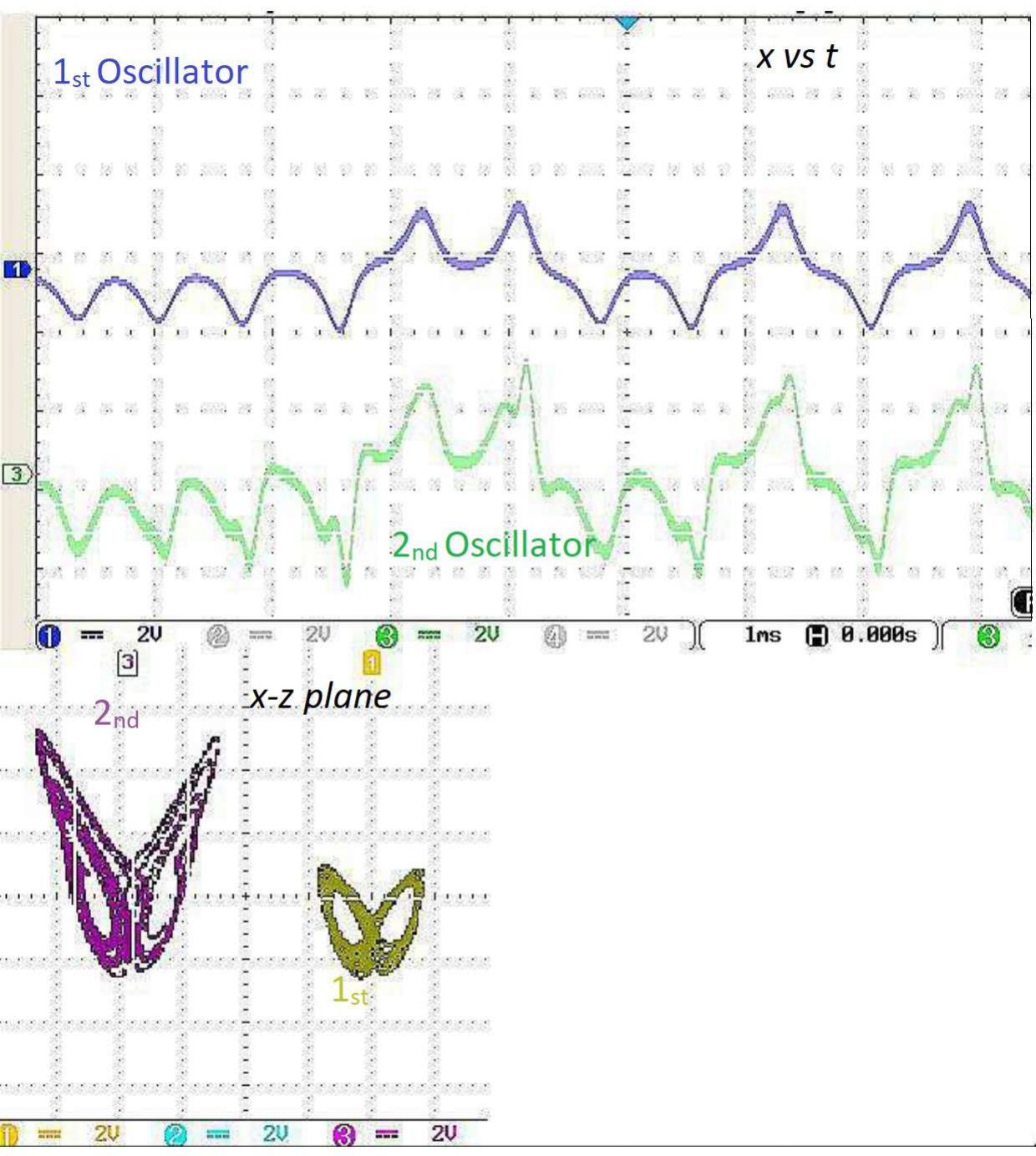}
\caption{Experimental results for the condition $y_1 = x_1$, $y_2 = x_2$, and $y_3 = \frac{x_3^2}{10}+\frac{9x_3}{10}$}
\label{fig:results2}
\end{figure}

The experimental results in Fig.~\ref{fig:results2} show the relationship between the signals \( x_1 \) and \( y_1 \) (blue and green, respectively), along with their phase space trajectories in the coupled oscillators (purple and yellow). To verify our results, we examine the \( x \)-waveform for both oscillators. Theoretically, these waveforms should be identical since \( y_1 = x_1 \), and the experimental results largely align with this expectation. Small fluctuations and nonsmooth transitions in the second waveform, likely due to noise, are observed. Despite these minor deviations, the overall behavior remains consistent with the expected dynamics.

To verify the relationship between \( y_3 \) and \( x_3 \), we observe the oscilloscope readings for the phase space diagram, where each square in the purple and yellow channels represents a voltage of 2 V. Substituting \( x_3 = 2 \, \text{V} \) into the equation
$y_3 = \frac{x_3^2}{10} + \frac{9x_3}{10},
$
we calculate \( y_3 = 2.2 \, \text{V} \). Additionally, in the phase space diagram of \( x \) vs. \( z \), the attractor for \( y_3 \) is observed to be approximately 2 to 2.5 times longer along the \( y \)-axis compared to that of the first oscillator. This aligns with the theoretical result, where \( y_3 \) is 2.2 times \( x_3 \), confirming that the specific relationship is satisfied.

\section{Conclusion}

In this work, we experimentally tested the generalized synchronization (GS) method by demonstrating projective synchronization and nonlinear scaling between coupled chaotic Lorenz oscillators. We specifically focused on unidirectional scaling, where the variables of one oscillator were scaled according to a predefined relationship with those of the other. The flexibility of our approach allows for easy extension to bidirectional coupling and exploration of other forms of relationships between the two oscillators.

\section*{Acknowledgement}

SB acknowledges the support from J C Bose National Fellowship provided by the Science and Engineering Research Board,
Government of India, Grant No. JBR/2020/000049.


\begin{thebibliography}{00}
\bibliographystyle{srt}

\bibitem{Kennedy_1992}
  M. Kennedy,  
  \textit{Robust OP Amp Realization of Chua's Circuit},  
  Frequenz,  
  vol. 46, no. 3-4, pp. 66-80, 1992. 

  \bibitem{lorenz1963}
  Edward N. Lorenz,
  \textit{Deterministic Nonperiodic Flow},
  Journal of the Atmospheric Sciences,
  vol. 20, no. 2, pp. 130--141, 1963,
  American Meteorological Society.

\bibitem{Rossler_1976}
  Otto Rössler,
  \textit{An equation for continuous chaos},
  Physics Letters A,
  Vol. 57, pp. 397-398,
  1976.

  \bibitem{Pecora_1990}
  Louis Pecora and T. Carroll,
  \textit{Synchronization in chaotic systems},
  Physical Review Letters,
  Vol. 64, p. 821,
  1990.

\bibitem{Fujisaka_1983}
  H. Fujisaka and T. Yamada,
  \textit{Stability theory of synchronized motion in coupled-oscillator systems},
  Progress of Theoretical Physics,
  Vol. 69, pp. 32-47,
  1983.
\bibitem{Pecora_1997}
  Louis M. Pecora, Thomas L. Carroll, Gregg A. Johnson, Douglas J. Mar, and James F. Heagy,
  \textit{Fundamentals of synchronization in chaotic systems, concepts, and applications},
  Chaos,
  Vol. 7, No. 4, pp. 520-543,
  1997.

\bibitem{Duolan_2023}
  Duolan Xiang, Linying Xiang, and Guanrong Chen,
  \textit{Synchronization of stochastic complex networks with time-delayed coupling},
  Chinese Physics B,
  Vol. 32, No. 6, p. 060502,
  2023.

  \bibitem{CHEN_2005}
  Hsien-Keng Chen,
  \textit{Synchronization of two different chaotic systems: A new system and each of the dynamical systems Lorenz, Chen, and Lü},
  Chaos, Solitons \& Fractals,
  Vol. 25, No. 5, pp. 1049-1056,
  2005.
  \bibitem{LU_2004}
  Jinhu Lü, Xinghuo Yu, and Guanrong Chen,  
  \textit{Chaos synchronization of general complex dynamical networks},  
  Physica A: Statistical Mechanics and its Applications,  
  vol. 334, no. 1, pp. 281-302, 2004. 
  \bibitem{Lin_2016}
  Xiaoran Lin, Shangbo Zhou, and Hua Li,
  \textit{Chaos and synchronization in complex fractional-order Chua’s system},
  International Journal of Bifurcation and Chaos,
  Vol. 26, No. 3, p. 1650046,
  2016.

\bibitem{Irimiciuc_2015}
  S.-A. Irimiciuc, O. Vasilovici, and D. G. Dimitriu,
  \textit{Chua’s circuit: Control and synchronization},
  International Journal of Bifurcation and Chaos,
  Vol. 25, No. 4, p. 1550050,
  2015.

\bibitem{Sun2009}
  Jie Sun, Erik M. Bollt, and Takashi Nishikawa,
  \textit{Constructing Generalized Synchronization Manifolds by Manifold Equation},
  SIAM Journal on Applied Dynamical Systems,
  Vol. 8, No. 1,
  pp. 202--221,
  2009.
  \bibitem{Rulkov_1995}
  Nikolai F. Rulkov, Mikhail M. Sushchik, Lev S. Tsimring, and Henry D. I. Abarbanel,
  \textit{Generalized synchronization of chaos in directionally coupled chaotic systems},
  Phys. Rev. E,
  vol. 51, no. 2, pp. 980--994, Feb 1995,
  American Physical Society.

  

\bibitem{KITTEL_1998}
  A. Kittel, J. Parisi, and K. Pyragas,
  \textit{Generalized synchronization of chaos in electronic circuit experiments},
  Physica D: Nonlinear Phenomena,
  Vol. 112, No. 3, pp. 459-471,
  1998.
\bibitem{Uchida_2003}
  Atsushi Uchida, Ryan McAllister, Riccardo Meucci, and Rajarshi Roy,  
  \textit{Generalized Synchronization of Chaos in Identical Systems with Hidden Degrees of Freedom},  
  Physical Review Letters,  
  vol. 91, no. 17, p. 174101, Oct. 2003.  

\bibitem{Abarbanel_1996}
  Henry D. I. Abarbanel, Nikolai F. Rulkov, and Mikhail M. Sushchik,  
  \textit{Generalized synchronization of chaos: The auxiliary system approach},  
  Physical Review E,  
  vol. 53, no. 5, pp. 4528-4535, May 1996.  

\bibitem{Parlitz_1996}
  U. Parlitz, L. Junge, W. Lauterborn, and L. Kocarev,  
  \textit{Experimental observation of phase synchronization},  
  Physical Review E,  
  vol. 54, no. 2, pp. 2115-2117, Aug. 1996. 
  
  \bibitem{Dmitriev_2009}
  Boris S. Dmitriev, Alexander E. Hramov, Alexey A. Koronovskii, Andrey V. Starodubov,  
  Dmitriy I. Trubetskov, and Yurii D. Zharkov,  
  \textit{First Experimental Observation of Generalized Synchronization Phenomena in Microwave Oscillators},  
  Physical Review Letters,  
  vol. 102, no. 7, p. 074101, Feb. 2009.
\bibitem{Ramaswamy_2018}
  Shahrukh Chishti and Ram Ramaswamy,
  \textit{Design strategies for generalized synchronization},
  Phys. Rev. E,
  vol. 98, no. 3, pp. 032217, Sep 2018,
  American Physical Society.


\bibitem{Koronovskii2014}
  A. A. Koronovskii, O. I. Moskalenko, A. S. Pavlov, N. S. Frolov, and A. E. Hramov,
  \textit{Generalized synchronization in the action of a chaotic signal on a periodic system},
  Technical Physics,
  vol. 59, no. 5, pp. 629--636, 2014.



\bibitem{Barajas_2012}
  J.G. Barajas Ramírez, K.P. Cuéllar Galarza, and R. Femat,
  \textit{Generalized synchronization of strictly different systems: Partial-state synchrony},
  Chaos, Solitons \& Fractals,
  Vol. 45, No. 3, pp. 193-204,
  2012.



\bibitem{Sun_2009}
  Jie Sun, Erik M. Bollt, and Takashi Nishikawa,
  \textit{Constructing generalized synchronization manifolds by manifold equation},
  SIAM Journal on Applied Dynamical Systems,
  Vol. 8, No. 1, pp. 202-221,
  2009.


  

 

\end{thebibliography}
\end{document}